\documentclass[a4paper]{article}
\usepackage{xcolor}
\usepackage{graphicx}
\graphicspath{ {./figures/} }
\usepackage{INTERSPEECH2020}
\usepackage{multirow}
\title{Streaming Language Identification using Combination of Acoustic Representations and ASR Hypotheses}
\name{Chander Chandak,     Zeynab Raeesy, Ariya Rastrow\\
Yuzong Liu, Xiangyang Huang, Siyu Wang,  Dong Kwon Joo,  Roland Maas}
\address{
  Amazon Alexa}
\email{\{cchach,raeesyzr,arastrow\}@amazon.com}

\begin{document}

\maketitle
\begin{abstract}
This paper presents our modeling and architecture approaches for building a highly accurate low-latency language identification system to support multilingual spoken queries for voice assistants. A common approach to solve multilingual speech recognition is to run multiple monolingual ASR systems in parallel and rely on a language identification (LID) component that detects the input language. Conventionally, LID relies on acoustic only information to detect input language. We propose an approach that learns and combines acoustic level representations with embeddings estimated on ASR hypotheses resulting in up to 50\% relative reduction of identification error rate, compared to a model that uses acoustic only features. Furthermore, to reduce the processing cost and latency, we exploit a streaming architecture to identify the spoken language early when the system reaches a predetermined confidence level, alleviating the need to run multiple ASR systems until the end of input query. The combined acoustic and text LID, coupled with our proposed streaming runtime architecture, results in an average of 1500ms early identification for more than 50\% of utterances, with almost no degradation in accuracy. We also show improved results by adopting a semi-supervised learning (SSL) technique using the newly proposed model architecture as a teacher model.

 \end{abstract}

\noindent\textbf{Index Terms}: speech recognition, multilingual speech, language identification, dynamic language switching

\section{Introduction}
Many households around the world now use voice assistants, embodied in far-field smart speakers, for daily tasks. For multilingual households, a monolingual voice interface, allowing a user to speak only in a single preset language, hinders functionality and creates friction in the experience. To allow users to dynamically switch between a selected pair of (or more) languages, the backend automatic speech recognition (ASR) service runs multiple speech recognizers in parallel, in conjunction with a separate language identification (LID) module. Once the language of an input voice query is detected, the corresponding ASR hypothesis is passed to downstream components, to execute the rest of the stack. The choice to run multiple ASRs in parallel, as opposed to first detect language and then recognize speech, is mainly to process speech utterances in a streaming fashion, which is paramount in building low-latency voice assistants. Furthermore, once the LID detects an input language, the non-matching ASR system(s) can be terminated, thus saving cost and resources. In this view, it is critical to not only build a highly accurate LID, but one that can identify a correct language fast and before the end of user's voice query.

Conventionally, the task of LID for speech queries relies on acoustic-only features and does not leverage semantic and lexical features. A caveat of an acoustic only system is that it could perform poorly on accented speech, over indexing on accent rather than content, and also when dealing with language pairs which are highly mixed and code-switched, e.g. English and Hindi. In this paper, we propose an LID approach that combines acoustic-based features with textual information captured in 1-best hypotheses of competing ASRs. Similar to approaches of~\cite{rmaas2018combining,mallidi2018device} used for other speech tasks, i.e. end-of-utterance and device-directed utterance detection, our model consists of two separate encoding components, namely acoustic encoding, which converts input audio to a learned embedding, and a text-based embedding derived from the 1-best output of each ASR. The final model is jointly trained to combine acoustic and text embeddings using an LID loss function. To adhere to a streaming LID, when using both acoustic and text information, the proposed architecture requires each ASR system to periodically (e.g. every 600 milliseconds) send its partial hypotheses output to an LID module. This entails extra communication between individual ASR systems and LID component. An acoustic-only LID, on the other hand, processes audio frames separately, not needing to access ASR hypotheses outputs. We further investigate in a semi-supervised training approach to improve an acoustic-only LID (student model) using unsupervised audio data labeled by a text-based LID teacher model. This proposes a simplified alternative for when building an intertwined runtime LID with ASR poses architectural challenges.
\vspace{-1mm}
\section{Related Work}\label{related-work}
\vspace{-1mm}

The language detection problem has been traditionally addressed by training a classifier with $N$ classes on a set of inputs extracted from the acoustic domain. Early language identifiers used i-vectors as an embedding input to the LID classifier~\cite{martinez2011language}. With the advent of deep neural network approaches, the focus shifted towards encoding the acoustic information using a deep learning framework, e.g. x-vectors~\cite{snyder2018spoken} and bottleneck features~\cite{fer2015multilingual}. Although these fixed-size embeddings can be used to train any discriminative classifier, e.g. discriminatively-trained Gaussian classifier~\cite{snyder2018spoken}, most recent approaches lean towards direct utilization of probabilities from the softmax layer~\cite{lopez2014automatic,gonzalez2014automatic,gonzalez2015frame,lopez2016use}. In order to scale to a large set of languages, without the need to train a separate model for each target language group,~\cite{wan2019tuplemax} proposes a new loss function \emph{tuplemax}, where at training the model is optimized to detect the language among $N$ classes, taking into account the prior that the classification decision is restricted to a predetermined subset of languages at runtime.

The aforementioned approaches rely only on acoustic features, missing the discriminative information captured in the lexical and semantic level~\cite{sefara2016text, duvenhage2019short,morris2020automated}. Recently, ~\cite{wang2019signal} investigated approaches to combine ASR-based features such as confidence, acoustic model, language model scores with text LID score (LID trained on ASR text) and acoustic LID. They concluded that the signal combination model, using DNNs, is a better alternative to a model trained only on acoustic features.
Our proposed method and study in this paper is mostly related to~\cite{wang2019signal} but different in two main aspects of a) input signals used in training the final combined LID model, and b) reporting improvements with respect to both accuracy gains and how early the input language is identified (cost savings). In our approach, we train three separate encoding layers, one to convert acoustic frames into a fixed-size embedding (acoustic embedding) and two encoders to convert 1-best hypotheses from each ASR into text embeddings. All encoding layers are trained using an LID loss function. These pre-trained embeddings (acoustic and text) are then used as an input to our final fully-connected classifier.  At runtime, we use a streaming based mechanism, periodically relaying ASR 1-best hypotheses to our LID module, and querying the final LID classifier using an interval based policy (every $T$ seconds). We analyze how much we can save in non-target ASR’s decoding cost by making a faster decision through combination model without compromising on accuracy.

\vspace{-2mm}

\section{Methods}\label{methods}
\vspace{-1mm}

Section~\ref{acoustic-lid} describes the acoustic-only language classifier. We then describe the details of text-based LID that is trained using ASR hypotheses in section~\ref{text-lid}, and finally present the signal combination model, AcousText model in section~\ref{combination}.
\subsection{Acoustic-based Language Identification}\label{acoustic-lid}

\begin{figure}[t]
\centering
  \centerline{\includegraphics[width=8.0cm]{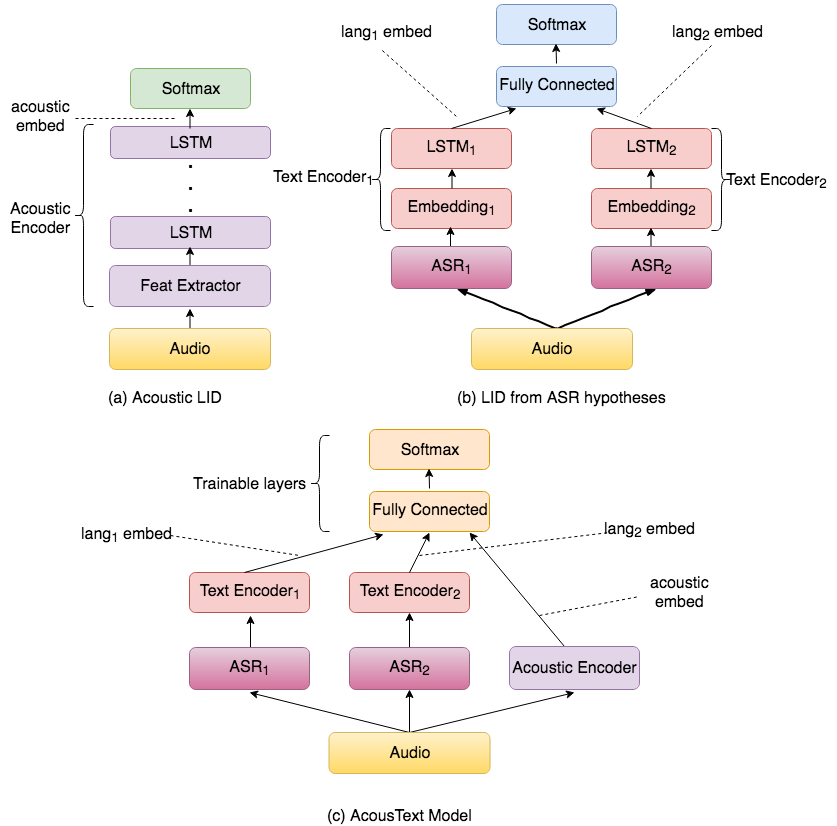}}
  \vspace{-1mm}
\caption{Schematic diagram for (a) acoustic-only LID,  (b) text-based LID that consumed ASR hypotheses from parallel ASR decoders, (c) acoustic + text-based LID }
\label{fig:design}
\end{figure}

Similar to ASR acoustic models, the acoustic-only LID (Figure~\ref{fig:design} (a)) consumes features extracted from audio, but the targets are language tags instead of speech units. For each frame a set of features are extracted and a target language label is defined. We used long short-term memory (LSTM) models because of their well-demonstrated ability to learn long-term dependencies in sequences.
The model can be trained on targets of all frames, a subset of frames, or alternatively on single utterance level target. Our observations suggest that training on every frame leads to overfitting while training on utterance-level target does not provide sufficient signal and yields an underfit model. We found the best approach to be backpropagating every $N$ frames. The training is done in two stages, first by chunking audio to smaller 36-frame sequences (with overlaps) to save training time followed by a few final epochs of training on full sequence of frames in the utterance.


The Acoustic LID is agnostic to intra-language speech unit differences and learns the language by learning the pronunciation, co-articulation, and sequences of phonemes that are more likely in each language. As a result, pure acoustic LID systems can misclassify an utterance spoken in language A with a strong accent of language B. Another challenging scenario for acoustic-only classifier is in acoustically similar languages where scripts and linguistics structure are the main distinctive features, such as Hindi (hi-IN) and Indian English (en-IN). Using some additional contextual information about the language is therefore necessary to address such issues. 

\subsection{Language Identification using ASR Hypotheses}\label{text-lid}
Our text-based LID (Figure~\ref{fig:design} (b)) is designed to predict the language from the texts generated by different ASR models. If the user had selected $k$ out of $n$ languages, $k$ ASR models are run in parallel. For each language, the text-based LID has a separate text encoder which consists of a character embedding layer and an LSTM layer.

Each of $k$ hypotheses is passed to its corresponding encoder. A vector of zeros is fed to the remaining $n$-$k$ text encoders. The final hidden state of the last LSTM layer of $k$th text encoder is considered as the text embedding for the $k$th language. We concatenate these embedding for each of the $n$ languages. The concatenated embedding is passed to fully connected layers to generate the softmax for $n$ languages. We have $n$ = $k$ = 2 as shown in the diagram in Figure~\ref{fig:design} for simplicity.

\subsection{AcousText LID}\label{combination}
\vspace{-1mm}
The combined model (Figure~\ref{fig:design} (c)) is called AcousText LID model. The input to the combined model consists of acoustic embeddings which is the final hidden state of the last LSTM layer of the acoustic LID model, and $n$ text embeddings, one true one for each of the $k$ languages, while $n-k$ text embeddings are zero vectors. 
As explained in section~\ref{text-lid}, audio is processed by $i$th ASR to generate $i$th hypothesis which is then passed to $i$th text encoder. The last LSTM state of $i$th encoder forms the $i$th text embedding. The text encoders and acoustic encoders are not trained in this stage.

\subsection{Inference}\label{inference}
\vspace{-1mm}

At runtime, the LID output probabilities are compared against a threshold at intervals of time $T$. Figure 2 shows an example where the inference being made at $t = 2T$ . At each interval, the 1-best (partial) hypotheses from both ASR models are passed to the corresponding text encoders to generate the respective text embeddings. In parallel, the input audio is processed by the acoustic encoder, frame-by-frame, to generate the acoustic embedding. The real-time factor for processing audio is likely to be different for the acoustic encoder and the ASR systems. Usually the simple acoustic encoder processes the audio faster than the ASR systems can generate partial 1-best outputs. To address this gap, the language is inferred at a particular interval only when both acoustic embedding and text embeddings are available for that interval.
Once both acoustic embeddings and text embeddings for a specific interval are ready, they are passed to the AcousText LID model to generate the posterior. If the posterior is greater than a tuned threshold $\theta$ (tuned on a development set), we conclude on the language for that utterance and terminate the competing ASR system. If the posterior is less than the threshold at all intervals until the end of input audio is reached, then the language is decided based on the maximum of the posteriors at the last interval.

\vspace{-2mm}
\section{Experiments}\label{experiments}
Our data studies across regions suggest that bilingual speakers form the majority of our multilingual speakers. Based on this and for simplifying the analysis, we perform our experiments with $N$ = 2 for three different language pairs: American English (en-US) vs American Spanish (es-US), Canadian English (en-CA) vs Canadian French (fr-CA), and Hindi (hi-IN) vs Indian English (en-IN). Most of the utterances in en-IN/hi-IN bucket are code-switched, where speakers switch between the two language within one utterance. We assign one language identity to the entire utterance. To account for utterances containing code-switching, we might have to wait until the end of the utterance to be able to make the correct decision. For the pairs without code-switching, we don't always need to wait until the end of the utterance and can predict the language with high confidence earlier.

For training our acoustic model, we use 64-dimensional log mel-filterbank (LFBE) feature vectors, calculated on 25ms windows with 10ms overlap. The utterance's language identity is used to set frame-wise labels (i.e. identical labels for all frames), and the model is trained by backpropagating only every 10 frames. Our network comprises of 3 unidirectional LSTM layers each with 768 cells. The training dataset ranged between 3k and 3.5k hours for each pair and was augmented by adding varying degrees of reverberation to clean audio (using room simulator).

For training our text-based model, for each utterance, two ASR hypotheses are generated by passing the audio to ASRs of both languages. For example, an es-US utterance is decoded using both en-US and es-US ASR models to generate the training data. For simplicity, we just take the final 1-best hypothesis from each ASR model. The text-based model contains an embedding layer of 128 units and one LSTM layer of 256 cells for each language as shown in Figure~\ref{fig:design}(b). The final state of the last LSTM layer forms the text embedding for that language. The embeddings are then passed to fully connected layers of 32 units and 2 output units before taking the softmax to generate the posteriors for each language. Dropout rate of 0.5 is applied after the LSTM and the fully connected layer. The text-based model is trained using around 1k hrs of audio for each language.

For training the AcousText model, acoustic embeddings are generated using the acoustic-only LID model and  text embeddings are generated using the text encoders described earlier. All three embeddings, one from the acoustic LID (with a dropout of 0.5) and two from text-based LIDs corresponding to each language, are then passed to fully connected layers of 64, 32 and 2 units before taking the softmax. Dropout rate of 0.5 is applied after each fully connected layer. The training data is generated in a similar fashion to text-based model (for simplicity the embeddings are only extracted from the final state). This model is trained with 350hrs of audio for each language. We use random anonymized interactions with Alexa for training and evaluating our models.


The inference module uses interval of T = 600ms in our experiments. 
Early stopping of the non-relevant ASR models saves computation cost. Table~\ref{table:savings} shows the results with early stopping for en-US/es-US and en-CA/fr-CA.
During inference, we choose intermediate 1-best hypothesis to generate the text embeddings but at training time we use the final 1-best hypotheses from both ASR models for a given utterance.

\begin{table*}[t]
\begin{center}
\vspace{-1mm}

\caption{\%Relative error rate reduction (RERR) calculated per language}\label{table:accuracy}
\vspace{-1mm}

\begin{tabular}{ccccccc}
\hline
\hline
Model & \multicolumn{2}{c}{English/Spanish} & \multicolumn{2}{c}{English/French} & \multicolumn{2}{c}{English/Hindi} \\
 \hline
 	    & en-US & es-US & en-CA & fr-CA & en-IN & hi-IN \\
\hline
Acoustic-only LID                       & 0.0 & 0.0 & 0.0 & 0.0 & 0.0 & 0.0\\
\hline
Text-based LID                   & 55.1 & 19.0 & 45.9 & 7.19 & 67.4 & 32.3\\
\hline
AcousText LID & 61.9 & 51.3 & 51.4 & 24.2 & 70.6 & 46.8\\
\hline 
\hline
\end{tabular}
\end{center}
\end{table*}
\vspace{-1mm}

\begin{table*}[t]
\begin{center}
\caption{\% Relative Error Rate Reduction (RERR), relative to acoustic-LID, percentage of decoding time saved (i.e. audio length that need not be processed further by ASR), and percentage of utterances where we detect the language early w.r.t threshold}\label{table:savings}
\begin{tabular}{ccccccccccccc}
\hline
\hline
 $\theta$ & \multicolumn{3}{c}{en-US} & \multicolumn{3}{c}{es-US} & \multicolumn{3}{c}{en-CA} & \multicolumn{3}{c}{fr-CA}\\
\hline
& \%RERR & \%saved & \%utt & \%RERR &  \%saved & \%utt &  \%RERR &   \%saved & \%utt & \%RERR &  \%saved & \%utt \\
\hline
0.99 & 58.8 & 43 & 56 & 37.7 & 40 & 65 & 43.4 & 46 & 58 & 18.2 & 38 & 54\\
\hline
0.95 & 50.1 & 46 & 72 & 24.2 & 48 & 83 & 41.4 & 48 & 74 & 13.6 & 45 & 70\\
\hline
\hline
\end{tabular}
\end{center}
\end{table*}



\begin{figure}[t!]
\centering
  \centerline{\includegraphics[width=8.0cm]{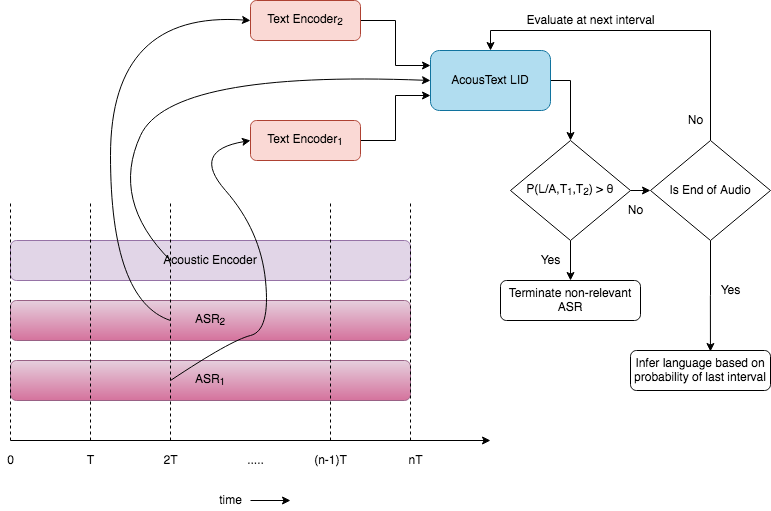}}
\caption{Schematic view of inference module at runtime.}
\label{fig:arbitration}
\end{figure}

\begin{table}[]
\begin{center}
\caption{Examples of ASR hypotheses generated by competing ASR systems (bold indicates intended language)}
\vspace{-1mm}

\label{table:examples}
\begin{tabular}{ccl}
\hline
\hline
\multirow{2}{*}{E1}            & en-IN & \footnotesize alexa summer of sixty nine songs play karo \\ \cline{2-3} 
                                          & hi-IN & \footnotesize  \textbf{alexa summer of sixty nine songs play karo}  \\ \hline         
\multirow{2}{*}{E2}                       & en-IN & \footnotesize \textbf{alexa  play chalo bulawa aaya hai naren chanchal} \\ \cline{2-3} 
                                          & hi-IN &  \footnotesize alexa  play chalo bulawa aaya hai naren chanchal  \\ \hline            
\multirow{2}{*}{E3}                       & en-CA &  \footnotesize  \textbf{who's the singer of Sous le ciel de Paris} \\ \cline{2-3} 
                                          & fr-CA &  \footnotesize  who's the singer of Sous le ciel de Paris  \\ \hline                                                                             
\multirow{2}{*}{E4}                       & en-US & alexa my notification \\ \cline{2-3} 
                                          & es-US &  \footnotesize  \textbf{alexa me notifica aquí}  \\ \hline 
\hline                                            
\end{tabular}
\end{center}
\end{table}

\section{Discussion}\label{discussion}

In section ~\ref{accuracy} we review the results in terms of error rate reduction and present some examples of special cases and note worthy observations. We conclude the section in~\ref{earlystopping} by a discussion around early stopping. 
\vspace{-1mm}

\subsection{Error Rate Reduction}\label{accuracy}
\vspace{-1mm}

Table~\ref{table:accuracy} presents the relative error rate reduction results on each language test set for acoustic-only, text-based, and AcousText LID models. The results show that using text information reduces error rate significantly (55\% and 51\% for en-US and en-CA). This might be because of more number of accented utterances in the test set on which acoustic-only LID model fails.

For the en-IN/hi-IN language pair, using text-based models reduces the error rate on both languages (67\%  and 46\% en-IN and hi-IN respectively). Although accent is not a major problem for this pair, code-switching can be a challenging issue. For example, in the command and control domain, a user can articulate the intent career phrase in English such as ``play'' while requesting a Hindi song title. The above caveats can successfully be addressed by incorporating hypotheses from ASR. The language of example $E1$ in Table~\ref{table:examples} can be determined only when the last word ``karo'' is processed. It contains an English entity ``summer of sixty nine songs" and the command (play karo) is code-switched with the English word ``play''and Hindi word ``karo". In this example, if we rely on acoustic-only LID and infer the language identity at any intervals before the last word is seen could results in selecting the incorrect language (en-IN). In example $E2$ of Table~\ref{table:examples}, we see that the entity name is a long Hindi word but the utterance is in English. Acoustic-only LID model prediction is incorrect because of the long trailing Hindi context. This kind of error can happen in English/French as shown in $E3$. Although both ASR models generate the same hypothesis in the examples $E1$, $E2$, and $E3$, identifying the correct language is necessary for downstream applications (e.g. TTS). 

On the other hand, text-based models can also have its caveats, depending on the design of ASR system. A Spanish utterance passed to an English ASR system might produce meaningful English text or Spanish ASR system can be designed to be partly bilingual and cover basic usages of English, hence producing identical hypothesis to the English ASR system. Example $E4$ in Table~\ref{table:examples} shows that the en-US ASR model generates an English hypothesis ``alexa my notification" for a Spanish utterance that has ``alexa me notifica aquí". Using text-based LID system would probably result in detecting the incorrect language. The proposed approach of combining acoustic and text information addresses the caveats in each system using the complementary input signals received from the other system. Although, it can be assumed that context information are implicitly provided in the audio, our experiments suggest that providing explicit text information that is representative of the linguistic patterns of the language is indeed useful. The reason might be that the text information acts as  compressed information from the audio that boosts the accuracy of the LID. 
\vspace{-1mm}

\subsection{Early Stopping}\label{earlystopping}
\vspace{-1mm}

The second metric we consider in this approach is the computational savings made by early detection of the correct language, that would allow for early termination of non-matching ASR. For Table~\ref{table:accuracy}, no early stopping was used, i.e. the entire audio was always decoded. Table~\ref{table:savings} shows that at threshold $\theta=0.99$ we are able to reduce the decoding of non-relevant ASR model by 43\% in case of English utterances and 40\% in case of Spanish utterances without a significant drop in accuracy. The percentage saved is the ratio of the length of audio in the test set that is processed only by a single ASR model (because the other ASR model is terminated by early decision) to the total length of audio where we terminated early expressed as percentage. \%utt is the percentage of utterances where we terminated early. We find similar computational savings in case of en-CA and fr-CA. As discussed earlier, we do not consider early stopping for language where code-switching is expected to happen often in the utterances.
\vspace{-1mm}

\subsection{Semi-Supervised Learning}
\vspace{-1mm}
The improvements from utilizing text-based information come with added runtime complexity as LID model relies on intermittent decoder hypotheses at each interval. Ideally, we strive to have the same high accuracy without the added complexity.

The baseline acoustic-only model was trained on randomly selected, anonymized, audio from monolingual stacks, hence an under-representation of target multilingual speakers. Speech data from the two groups are different in two main attributes: accent (Latin pairs) and code-switching (Hindi/English). We used semi-supervised learning to get better representation of multilingual end-users' speech in training.

%
%
%
%

We select 1 million anonymized live traffic streams (1,100 hours) from multilingual speakers (who activated bilingual mode on their devices). Next, we ran the streams through text-based LID and discarded streams with posterior probabilities $<$ 99\%. The remaining data with labels from text-based LID were added to the original datasets to retrain the acoustic-LID.

\begin{table}[t]
\begin{center}
\caption{\%Relative Error Rate Reduction of the acoustic-only LID model retrained after data augmentation}\label{table:SSL}
\vspace{-1mm}

\begin{tabular}{ccccc}
\hline
\hline
 Model & \multicolumn{2}{c}{English/Spanish} & \multicolumn{2}{c}{English/Hindi} \\
 \hline
 	    & en-US & es-US & en-IN & hi-IN \\
\hline
baseline acoustic-only          &    0.0 & 0.0 & 0.0 & 0.0 \\
\hline
+ SSL          & 48.4 & 14.71 & 20.0 &17.6 \\
\hline 
\hline
\end{tabular}
\end{center}
\end{table}

Table~\ref{table:SSL} shows acoustic-only LID models trained with added SSL data outperform the baseline model through better matching train and target application. We hypothesize the higher gains in en-US are due to higher adoption by native Spanish speakers. For Hindi/English, this approach helps because it corrects the labels based on underlying linguistic pattern. 


\vspace{-2mm}

\section{Conclusion}
\vspace{-1mm}

We demonstrated how the combined text and acoustic embeddings boosts the accuracy of the language identification systems by overcoming the caveats of acoustic-only and text-based systems. In addition, we showed how the accuracy gap between acoustic-only and combined model can be reduced by semi-supervised learning, using text-based LID as a teacher model. 

\bibliographystyle{IEEEtran}

\bibliography{mybib}

\begin{thebibliography}{10}
\providecommand{\url}[1]{#1}
\csname url@samestyle\endcsname
\providecommand{\newblock}{\relax}
\providecommand{\bibinfo}[2]{#2}
\providecommand{\BIBentrySTDinterwordspacing}{\spaceskip=0pt\relax}
\providecommand{\BIBentryALTinterwordstretchfactor}{4}
\providecommand{\BIBentryALTinterwordspacing}{\spaceskip=\fontdimen2\font plus
\BIBentryALTinterwordstretchfactor\fontdimen3\font minus
  \fontdimen4\font\relax}
\providecommand{\BIBforeignlanguage}[2]{{%
\expandafter\ifx\csname l@#1\endcsname\relax
\typeout{** WARNING: IEEEtran.bst: No hyphenation pattern has been}%
\typeout{** loaded for the language `#1'. Using the pattern for}%
\typeout{** the default language instead.}%
\else
\language=\csname l@#1\endcsname
\fi
#2}}
\providecommand{\BIBdecl}{\relax}
\BIBdecl

\bibitem{rmaas2018combining}
R.~Maas, A.~Rastrow, C.~Ma, G.~Lan, K.~Goehner, G.~Tiwari, S.~Joseph, and
  B.~Hoffmeister, ``Combining acoustic embeddings and decoding features for
  end-of-utterance detection in real-time far-field speech recognition
  systems,'' in \emph{2018 IEEE International Conference on Acoustics, Speech
  and Signal Processing (ICASSP)}.\hskip 1em plus 0.5em minus 0.4em\relax IEEE,
  2018, pp. 5544--5548.

\bibitem{mallidi2018device}
S.~H. Mallidi, R.~Maas, K.~Goehner, A.~Rastrow, S.~Matsoukas, and
  B.~Hoffmeister, ``Device-directed utterance detection,'' \emph{arXiv preprint
  arXiv:1808.02504}, 2018.

\bibitem{martinez2011language}
D.~Martinez, O.~Plchot, L.~Burget, O.~Glembek, and P.~Mat{\v{e}}jka, ``Language
  recognition in ivectors space,'' in \emph{Twelfth annual conference of the
  international speech communication association}, 2011.

\bibitem{snyder2018spoken}
D.~Snyder, D.~Garcia-Romero, A.~McCree, G.~Sell, D.~Povey, and S.~Khudanpur,
  ``Spoken language recognition using x-vectors.'' in \emph{Odyssey}, 2018, pp.
  105--111.

\bibitem{fer2015multilingual}
R.~F{\'e}r, P.~Mat{\v{e}}jka, F.~Gr{\'e}zl, O.~Plchot, and
  J.~{\v{C}}ernock{\`y}, ``Multilingual bottleneck features for language
  recognition,'' in \emph{Sixteenth Annual Conference of the International
  Speech Communication Association}, 2015.

\bibitem{lopez2014automatic}
I.~Lopez-Moreno, J.~Gonzalez-Dominguez, O.~Plchot, D.~Martinez,
  J.~Gonzalez-Rodriguez, and P.~Moreno, ``Automatic language identification
  using deep neural networks,'' in \emph{2014 IEEE international conference on
  acoustics, speech and signal processing (ICASSP)}.\hskip 1em plus 0.5em minus
  0.4em\relax IEEE, 2014, pp. 5337--5341.

\bibitem{gonzalez2014automatic}
J.~Gonzalez-Dominguez, I.~Lopez-Moreno, H.~Sak, J.~Gonzalez-Rodriguez, and
  P.~J. Moreno, ``Automatic language identification using long short-term
  memory recurrent neural networks,'' in \emph{Fifteenth Annual Conference of
  the International Speech Communication Association}, 2014.

\bibitem{gonzalez2015frame}
J.~Gonzalez-Dominguez, I.~Lopez-Moreno, P.~J. Moreno, and
  J.~Gonzalez-Rodriguez, ``Frame-by-frame language identification in short
  utterances using deep neural networks,'' \emph{Neural Networks}, vol.~64, pp.
  49--58, 2015.

\bibitem{lopez2016use}
I.~Lopez-Moreno, J.~Gonzalez-Dominguez, D.~Martinez, O.~Plchot,
  J.~Gonzalez-Rodriguez, and P.~J. Moreno, ``On the use of deep feedforward
  neural networks for automatic language identification,'' \emph{Computer
  Speech \& Language}, vol.~40, pp. 46--59, 2016.

\bibitem{wan2019tuplemax}
L.~Wan, P.~Sridhar, Y.~Yu, Q.~Wang, and I.~L. Moreno, ``Tuplemax loss for
  language identification,'' in \emph{ICASSP 2019-2019 IEEE International
  Conference on Acoustics, Speech and Signal Processing (ICASSP)}.\hskip 1em
  plus 0.5em minus 0.4em\relax IEEE, 2019, pp. 5976--5980.

\bibitem{sefara2016text}
T.~J. Sefara, M.~J. Manamela, and P.~T. Malatji, ``Text-based language
  identification for some of the under-resourced languages of south africa,''
  in \emph{2016 International Conference on Advances in Computing and
  Communication Engineering (ICACCE)}.\hskip 1em plus 0.5em minus 0.4em\relax
  IEEE, 2016, pp. 303--307.

\bibitem{duvenhage2019short}
B.~Duvenhage, ``Short text language identification for under resourced
  languages,'' \emph{arXiv preprint arXiv:1911.07555}, 2019.

\bibitem{morris2020automated}
V.~Morris, ``Automated language identification of bibliographic resources,''
  \emph{Cataloging \& Classification Quarterly}, vol.~58, no.~1, pp. 1--27,
  2020.

\bibitem{wang2019signal}
S.~Wang, L.~Wan, Y.~Yu, and I.~L. Moreno, ``Signal combination for language
  identification,'' \emph{arXiv preprint arXiv:1910.09687}, 2019.

\end{thebibliography}

\end{document}